\documentclass[conference,compsoc]{IEEEtran}
\ifCLASSOPTIONcompsoc
  \usepackage[nocompress]{cite}
\else
  \usepackage{cite}
\fi
\ifCLASSINFOpdf
  \usepackage[pdftex]{graphicx}
\else
  \usepackage[dvips]{graphicx}
\fi
\usepackage{xcolor}
\usepackage{etoolbox}

\usepackage{xspace}
\newcommand{\cellmate}{%
  \textsc{ce}\scalebox{1.3}{\textsc{llm}}\textsc{ate}\xspace%
}
\newcommand{\cellmates}{\cellmate's\xspace}

\usepackage{listings}

\lstdefinestyle{common}{
  basicstyle=\ttfamily\small,
  columns=fullflexible,
  keepspaces=true,
  breaklines=true,
  breakatwhitespace=true,
  frame=single,
  framerule=0.4pt,
  showstringspaces=false,
  tabsize=2,
  numbers=left,
  numberstyle=\tiny,
  numbersep=8pt,
  captionpos=b
}

\lstdefinelanguage{JavaScript}{
  keywords={break, case, catch, class, const, continue, debugger, default,
    delete, do, else, export, extends, finally, for, function, if, import,
    in, instanceof, let, new, return, super, switch, this, throw, try, typeof,
    var, void, while, with, yield, async, await},
  keywordstyle=\color{blue},
  ndkeywords={true, false, null, undefined},
  ndkeywordstyle=\color{purple},
  sensitive=true,
  comment=[l]{//},
  morecomment=[s]{/*}{*/},
  commentstyle=\color{gray},
  stringstyle=\color{red},
  morestring=[b]',
  morestring=[b]"
}

\definecolor{eclipseGreen}{RGB}{0, 128, 1}
\definecolor{eclipseRed}{RGB}{165, 42, 42}
\definecolor{eclipseGray}{RGB}{128, 128, 128}
\definecolor{eclipsePurple}{RGB}{158, 49, 246}

\newcommand{\jsonkey}[1]{{\color{eclipseGreen}\textbf{#1}}}

\lstdefinestyle{jsonStyle}{
    basicstyle=\ttfamily\scriptsize,   
    columns=flexible,
    keepspaces=true,
    frame=single,                 
    numbers=left,                 
    numberstyle=\tiny\color{black},
    numbersep=10pt,
    xleftmargin=0.02\linewidth,
    linewidth=\columnwidth,       
    breaklines=true,
    breakatwhitespace=false,
    breakindent=24pt, 
    breakautoindent=true,
    postbreak=\mbox{\textcolor{gray}{$\hookrightarrow$}\space},
    captionpos=b,
    stringstyle=\color{eclipseRed},
    morestring=[b]",
    keepspaces=true,
    showstringspaces=false,
    literate=
     {"semantic_action"}{{{\jsonkey{"semantic\_action"}}}}{1}
     {"description"}{{{\jsonkey{"description"}}}}{1}
     {"url"}{{{\jsonkey{"url"}}}}{1}
     {"method"}{{{\jsonkey{"method"}}}}{1}
     {"body"}{{{\jsonkey{"body"}}}}{1}
     {"args"}{{{\jsonkey{"args"}}}}{1}
     {"type"}{{{\jsonkey{"type"}}}}{1}
     {"source"}{{{\jsonkey{"source"}}}}{1}
     {"selector"}{{{\jsonkey{"selector"}}}}{1}
     {"name"}{{{\jsonkey{"name"}}}}{1}
     {"effect"}{{{\jsonkey{"effect"}}}}{1}
     {"actions"}{{{\jsonkey{"actions"}}}}{1}
     {"condition":}{{{\jsonkey{"condition":}}}}{1}
     {"parameters"}{{{\jsonkey{"parameters"}}}}{1}
     {"domain"}{{{\jsonkey{"domain"}}}}{1}
     {"selected_policies"}{{{\jsonkey{"selected\_policies"}}}}{1}
     {"allowed_domains"}{{{\jsonkey{"allowed\_domains"}}}}{1}
     {"task":}{{{\jsonkey{"task":}}}}{1}
     {"ground_truth"}{{{\jsonkey{"ground\_truth"}}}}{1}
     {"prediction"}{{{\jsonkey{"prediction"}}}}{1}
     {"ground_truth_args":}{{{\jsonkey{"ground\_truth\_args":}}}}{1}
     {"predicted_args":}{{{\jsonkey{"predicted\_args":}}}}{1}
     {"add_to_cart":}{{{\jsonkey{"add\_to\_cart":}}}}{1}
     {"update_quantity":}{{{\jsonkey{"update\_quantity":}}}}{1}
     {"view_shopping_cart":}{{{\jsonkey{"view\_shopping\_cart":}}}}{1}
     {"purchase_amount_leq":}{{{\jsonkey{"purchase\_amount\_leq":}}}}{1}
     {"quantity":}{{{\jsonkey{"quantity":}}}}{1}
     {"maxAmount"}{{{\jsonkey{"maxAmount"}}}}{1}
     {"totalAmount":}{{{\jsonkey{"totalAmount":}}}}{1}
     {"checkinDate"}{{{\jsonkey{"checkinDate"}}}}{1}
     {"checkoutDate"}{{{\jsonkey{"checkoutDate"}}}}{1}
     {"numGuests"}{{{\jsonkey{"numGuests"}}}}{1}
}

\lstdefinestyle{jsStyle}{
    language=JavaScript,
    basicstyle=\ttfamily\scriptsize,
    columns=flexible,
    basewidth=0.5em,
    keepspaces=true,
    numbers=left,                 
    numberstyle=\tiny\color{black},
    numbersep=10pt,
    frame=single,
    xleftmargin=0.02\linewidth,
    linewidth=\columnwidth,
    breaklines=true,
    breakatwhitespace=false,
    breakautoindent=true,
    breakindent=24pt,
    postbreak=\mbox{\textcolor{gray}{$\hookrightarrow$}\space},
    stringstyle=\color{eclipseRed},
    commentstyle=\color{gray},
    classoffset=0,
    keywords={export, default, function, const, let, var, if, else, return, true, false, try, catch, switch, case, while, for},
    keywordstyle=\color{eclipseGreen}\bfseries,
    classoffset=1,
    keywords={typeof, new, this, void, instanceof, delete},
    keywordstyle=\color{eclipsePurple}\bfseries,
    classoffset=0,
    morestring=[b]",
    morestring=[b]',   
    literate={\ }{{\ }}1,
}

\usepackage{subcaption} 

\usepackage{booktabs}
\usepackage{multirow}
\usepackage{makecell} 
\usepackage{array}
\newcolumntype{P}[1]{>{\centering\arraybackslash}p{#1}}

\usepackage{xurl} 
\usepackage[hidelinks]{hyperref}

\begin{document}
%
\title{
  \Large \bf \cellmate: Sandboxing Browser AI Agents
}

\author{\IEEEauthorblockN{Luoxi Meng}
\IEEEauthorblockA{UC San Diego\\lumeng@ucsd.edu}
\and
\IEEEauthorblockN{Henry Feng}
\IEEEauthorblockA{UC San Diego\\hefeng@ucsd.edu}
\and
\IEEEauthorblockN{Ilia Shumailov}
\IEEEauthorblockA{AI Sequrity Company\\ilia@sequrity.ai}
\and
\IEEEauthorblockN{Earlence Fernandes}
\IEEEauthorblockA{UC San Diego\\efernandes@ucsd.edu}
}


%


\maketitle

\begin{abstract}
Browser-using agents (BUAs) are an emerging class of AI agents that interact with web browsers in human-like ways, including clicking, scrolling, filling forms, and navigating across pages. While these agents help automate repetitive online tasks, they are vulnerable to prompt injection attacks that trick an agent into performing undesired actions, such as leaking private information or issuing unintended state-changing requests. We propose \cellmate, a browser-level sandboxing framework that restricts the agent's ambient authority and reduces the blast radius of prompt injections. We address the semantic gap challenge that is fundamental to BUAs --- writing and enforcing security policies for low-level UI tools like clicks and keystrokes is brittle and error-prone.  Our core insight is to perform sandboxing at the HTTP layer because all side-effecting UI operations will result in network communication to the website's backend. We implement \cellmate as an agent-agnostic browser extension and demonstrate how it enables sandboxing policies that block prompt injection attacks in the WASP benchmark with 7.25--15\% latency overhead.
\end{abstract}


%
\IEEEpeerreviewmaketitle

\section{Introduction}

Browser-using agents (BUA) interact autonomously with websites to accomplish user-specified tasks. Examples include Gemini-CUA~\cite{gemini-cua}, OpenAI Atlas~\cite{openai-atlas}, Perplexity Comet~\cite{perplexity-comet}, Anthropic Claude-for-Chrome~\cite{anthropic_claude_chrome_docs}, Browser-Use~\cite{browser-use}. As input, they receive screenshots and/or HTML trees and manipulate the website in the same way a human would --- through UI actions like clicking, typing and scrolling. Despite the significant advantages these agents offer by automating routine tasks, they are vulnerable to prompt injection attacks that trick the agent into performing dangerous tasks such as leaking private information or issuing unintended state-changing requests. There are many documented instances of attacks that affect real systems and users~\cite{comet-pi,wunderwuzzi-openai-pi,rehberger2025aiclickfix}. Willison describes a combination of three factors that make agents ripe for exploitation: sensitive resource access, external communication, and untrusted content~\cite{willison2025lethaltrifecta}. BUAs support all three conditions, using only the set of UI observation and manipulation tools. 

This paper describes \cellmate, a sandboxing framework for BUAs that enforces deterministic guardrails outside the agent at the browser-level.  Our work composes with existing ML-based defenses that either train models to resist prompt injections~\cite{secalign,chen2025meta,wallace2024instruction,learnprompting2023sandwich,hines2024spotlighting,walter2025softdeescalation} or train models to detect them~\cite{liu2025datasentinel,hung2025attentiontrackerdetectingprompt}. By building our sandbox into the \textit{environment} that the BUA interacts with, we escape the perpetual arms race in adversarial machine learning where adaptive attackers appear to always break ML-based defenses~\cite{rando2025adversarialmlproblemsgetting,pandya2025may,athalye2018obfuscatedgradientsfalsesense,nasr2025attackermovessecondstronger}.



Creating a sandboxing framework for BUAs poses unique, fundamental challenges in both policy enforcement and specification. The first challenge is defining a meaningful sandbox boundary. Consider a policy such as: ``\textit{the agent may only check out on Amazon if the total amount is less than 50 dollars}''. Where should such a policy be enforced? Because the BUA interacts with the browser through a series of UI observations and manipulations (e.g., clicking, typing, scrolling), one might attempt to enforce policies directly at this level by constraining these low-level primitive actions.
However, this approach is brittle and error-prone: imagine a policy saying that the agent can click (246, 1023) but not other regions. It is non-sensical because those same co-ordinates can have very different semantics depending on the website, its state, screen resolution, etc. This is the \textit{semantic gap}, a long-standing challenge in computer security~\cite{sommer-paxson,bhushan-semantic-gap,brendan-semantic-gap}. Expressing the policy and enforcing it do not occur at the same level of abstraction. This problem is not as severe in traditional computer systems because of layered abstractions: the process-kernel system call interface or the Android app-to-framework interface exist at abstraction levels where semantically meaningful policies can be written and enforced.

The second challenge is about how policies are created. In a traditional system, the platform developer defines the policy/permission system (e.g., Linux access control lists at the level of files, Android permissions at the level of system services~\cite{android-perms}, OAuth scopes at the level of web APIs~\cite{hardt2025oauth2.1}). The app developer either selects from a set of existing policies/permissions (e.g., Android app developers select a best-match from the set of platform-defined permissions) or a system administrator configures a mandatory policy (e.g., SELinux~\cite{smalley2006implementing}). In the BUA setting,  there is no app or developer \textit{per se};  we only have a natural language description of the user's task.

Our sandboxing framework addresses these challenges in principled ways. At a high-level, \cellmate interposes on HTTP requests at the browser level and then evaluates a policy on the HTTP request and its parameters.  Our key insight is that even though BUAs interact with websites through low-level UI actions, those actions ultimately result in HTTP messages to the website's backend. This insight bridges the semantic gap because HTTP messages have inherent meaning (unlike clicks and keystrokes). 


A key question is, how does \cellmate know about which HTTP requests are salient for a policy? For our running example of purchasing an item on Amazon only if the amount is below a certain value, \cellmate needs to know which HTTP request to intercept and what parameters to use during policy evaluation. We introduce the concept of an \textit{agent sitemap} to address this issue. Just as a regular sitemap lists various URLs to assist web crawlers, the agent sitemap lists all HTTP messages. We expect that either the website developers or their security teams would create these agent sitemaps and host them at a well-known location (e.g., on the website domain itself). This expectation is grounded in established precedent: developers already create and maintain similar security-critical metadata including Content Security Policy headers, robots.txt files, and OAuth scopes. Agent sitemaps fit naturally into this ecosystem: like CSP, they protect end users; like robots.txt, they guide automated agents; like OAuth scopes, they define permission boundaries. Our hope is that the agent sitemap becomes a public standard that enables websites to comply with upcoming AI safety regulations such as the EU AI Safety Act.



A sandboxing framework relies on policies that define what actions are allowed.  A policy involves three aspects: (1) who creates it? (2) when is it created? (3) how is it created? There is a large design spectrum of policy architectures that needs to be explored to answer these questions~\cite{iqbal,camel-cua,fides,shi2025progent}.   \cellmate is independent of policy architecture --- it will enforce policies as long as they adhere to our agent sitemap guidelines. However, in the interest of prototyping an end-to-end system that can protect users today, we design and implement a discretionary policy mechanism on top of \cellmate.

We take inspiration from existing successful policy/permission designs in smartphone and web systems. Just like web developers today define OAuth permissions for REST APIs or the Android designers defined permissions for system resources, we expect that web app developers will define similar policies over the HTTP resources identified in the agent sitemap. For our running example of purchasing an item on Amazon, an example policy might be defined for the HTTP method that performs the checkout and will have semantics of ensuring that the purchase price is less than a configurable value. Based on this developer-provided universe of policies, the \cellmate policy architecture solves the problem of \textit{policy selection}: given a natural language user task, select and instantiate the policies that best match the user's current task.

\noindent\textbf{Contributions.}
\begin{itemize}
    \item We design the first systems-level sandboxing framework, \cellmate, for browser-using agents that requires cooperation from various stakeholders in the ecosystem: users, browsers, and web app developers. The user specifies natural language tasks, web app developers define agent sitemaps, and browsers enforce policies at the HTTP level. The various components of the framework align with the interests of each stakeholder. For instance, web app developers want to protect their users, so they are motivated and equipped to create agent sitemaps. 

    \item We instantiate \cellmate for Chrome by structuring it as a browser extension. Our design and implementation is agent-agnostic and can protect users independent of the specific BUA being used.

    \item We design and implement a policy layer on top of \cellmate to show its flexibility and applicability in protecting users today. We create a policy selection benchmark and characterize the ability of modern reasoning LLMs to automatically select and specialize policies for user tasks specified in natural language. Our evaluation shows that state-of-the-art LLMs can perform the policy selection and instantiation task with a high overall accuracy above 94\% across all task categories.
\end{itemize}

\cellmate is open-source at \texttt{\url{https://cellmate-sandbox.github.io}}.

\section{Background and Motivation}

\noindent\textbf{Browser-Using Agents (BUAs).} Browser-using agents (BUAs)~\cite{gemini-cua,openai-atlas,anthropic-cua,browser-use,perplexity-comet} represent a new class of autonomous agents that interact with web browsers in human-like ways, such as clicking, scrolling, filling forms, and navigating across multiple pages.
A BUA typically consists of a large language model (LLM) and a runtime environment. The runtime leverages browser automation frameworks such as Playwright~\cite{playwright} or Puppeteer~\cite{puppeteer} to observe browser state and interact with web applications.
At its core, a BUA operates in an iterative agent loop. In each iteration, the agent captures the state of the current browser tab (e.g., screenshot and DOM); then, it queries the LLM to select the next actions from a predefined set (e.g., click, scroll, navigate) with necessary arguments (URL or DOM element index) and execute them in the browser. This repeats until the task is complete. Uniquely, BUAs use low-level tools without inherent semantic meaning. For example, \texttt{click()} may mean read or delete an email depending on the target co-ordinate. By contrast, other agentic systems~\cite{qin2023toolllm, schick2023toolformer,patil2024gorilla} use higher-level tools with distinct semantics (e.g., send or delete an email.)

\noindent\textbf{Prompt Injection in BUAs.}
Prompt injection is a major security risk in LLMs, where maliciously crafted text can trigger unauthorized actions or data leaks, akin to traditional exploits like stack smashing.
BUAs inherit this vulnerability from their underlying LLMs. For example, Perplexity's Comet~\cite{perplexity-comet} forwards webpage content to its backbone LLM, allowing attackers to control user accounts~\cite{comet-pi}, and OpenAI's Operator~\cite{openai-operator} can be manipulated to leak private email addresses~\cite{wunderwuzzi-openai-pi}.
Operating with full access to authenticated browser sessions, BUAs combine ambient privilege with untrusted input, a \textit{lethal trifecta}~\cite{willison2025lethaltrifecta} that allows an agent to potentially take any action on behalf of a user.
The threat is further amplified by the growing ease of crafting prompt injections, both manually (e.g., linguistic quirks, role-playing) and automatically via optimization-based algorithms~\cite{gcg, paulus2024advprompter}.
Consequently, prompt injection poses an urgent challenge for BUAs.

\noindent\textbf{System-Level Defenses against Prompt Injection.}
In traditional computer security, system-level techniques such as access control and sandboxing have provided robust security by enforcing the \textit{principle of least privilege}~\cite{polp}: systems should operate with only the minimal permissions necessary to complete their tasks, thereby containing potential harm.
Inspired by this principle, our key insight is that security guarantees for LLM-integrated applications must be achieved at the \textit{system} level. 
Recent efforts have started to embrace this perspective. 
CaMeL~\cite{camel} and Fides~\cite{fides}, building on Willison's Dual LLM~\cite{willison2023dualllm}, separate trusted and untrusted contexts and enforce control and data flow derived from the trusted context. 
Progent~\cite{shi2025progent} restricts agents' behavior by enforcing privilege control policies on tool invocations.
However, all these approaches implicitly assume that policies can be enforced solely by restricting tool usage --- a premise that relies on a clear mapping between tool interfaces and security boundaries.
This assumption does not hold for BUA tools, where the effects of low-level actions (e.g., clicks or keystrokes) are dynamically determined by the execution context, including page state, DOM structure, and screen resolution.
As a result, translating meaningful security policies (e.g., ``do not delete emails from my Gmail inbox'') into constraints on these non-semantic primitives is inherently challenging.
Parallel to \cellmate, Foerster et al.~\cite{camel-cua} scale CaMeL to computer-use agents by adopting an observe-verify-act paradigm within execution plans. However, this approach suffers from the same limitation as its predecessor: it relies on the assumption of semantic tools. Furthermore, as discussed by Foerster et al., their system remains vulnerable to branch steering attacks where UI elements are manipulated to trigger unintended yet valid actions --- a vector that \cellmate explicitly neutralizes. 
 \cellmate is the first system-level defense that decouples policy enforcement from low-level tool interfaces.



\section{\cellmate Design}
\cellmate is a browser-level enforcement mechanism that supports flexible policy specification for agent-driven web interactions via precise, site-specific access control.
\cellmate enables expressive security policies; for instance, an agent is \textit{only allowed to comment on GitHub issues}, or \textit{can only place orders below a fixed amount on Amazon}.

\subsection{Threat Model and System Assumptions}
\label{sec:design-threat-model}


Our design goal is to create a framework for programmatic privilege control of BUAs. Our work sandboxes the action space of BUAs using policies that can be specified by multiple stakeholders (administrators, end-users, website developers, third-party interest groups). The policies are enforced inside the browser at the HTTP level.


\noindent\textbf{Attacker Goals and Capabilities.}
We assume prompt injection attackers who operate under realistic constraints. In particular, attackers may control untrusted portions of otherwise trusted domains (i.e., they do not control amazon.com or github.com, but may control the title and description of a GitHub issue or product reviews on Amazon).
Attackers can also control domains on the Internet that a user might mistakenly ask the agent to visit. Attackers can use this to launch attacks on trusted websites (e.g., via redirects and TOCTOU attacks~\cite{toctou-openai, jones2025systematization}). These assumptions reflect real attacks~\cite{comet-pi,wunderwuzzi-openai-pi,rehberger2025aiclickfix}. The attacker's goal is to exploit the ambient privilege of BUAs and violate the confidentiality and integrity of user data. Attackers can use a variety of prompt injection techniques, including optimization-based~\cite{pasquini2024neuralexec,gcg,fu2024imprompter,fun-tuning,pandya2025may} and optimization-free methods~\cite{tap,actor-critic-gemini,nasr2025attackermovessecondstronger}. 




\noindent\textbf{Non-Goals.}
\cellmate limits the action space of agents to what is necessary  for task completion, given an appropriate policy. Implementing data flow  or control flow integrity mechanisms (e.g., in the style of CaMeL~\cite{camel} or  Fides~\cite{fides}) is outside scope. For example, if an agent reads untrusted data from a webpage and later makes an HTTP request, \cellmate does not track  whether that request was causally influenced by the untrusted data --- it only  enforces whether the request itself is permitted by policy.


\noindent\textbf{Assumptions on Web Developers and User Tasks.}
\label{sec:design-assumptions}
Securing BUAs requires the co-ordinated effort of all stakeholders.
Developers of trusted domains are motivated to protect sensitive resources and their users’ sessions. Users are similarly motivated to protect their own information, but they often lack the security expertise to defend against subtle threats such as prompt injection attacks. Thus, we do not expect users to handle complex security configurations.
Since \cellmate is designed as a browser-level enforcement framework that supports various permission designs, certain assumptions may be specific to a particular permission system.
In Section~\ref{sec:design-permission}, we discuss one such design and its associated assumptions.

\noindent\textbf{System Assumptions.}
We assume a standard, trusted browser runtime. \cellmate requires a mechanism for intercepting HTTP traffic issued by the agent-controlled browser session. There are many implementation techniques to provide this property and we discuss the trade-offs in Section~\ref{sec:impl-enforcement-interception}. At a high-level, we choose to prototype \cellmate as a Chrome extension, and therefore, we must assume that no other malicious extensions are present that intercept/block HTTP requests.\footnote{If malicious extensions are present, then the user has bigger problems than prompt injection attackers.} In principle, \cellmate could be implemented as a core browser service and this will remove the assumption about the presence of other extensions, but this increases deployability issues unless the browser vendor itself implements the mechanism. The design principles in our work are independent of implementation choices.


Because \cellmate focuses on controlling agents' access to web applications, we assume that BUAs are confined to actions within the webpage context (e.g., clicking, typing, navigating), and cannot alter browser settings, access local files, or use developer tools.
In enterprises, which tend to be early adopters for browser agents, it is common practice for system administrators to define Chrome policies to enforce the necessary browser-level constraints~\cite{chrome-enterprise-policies}.




\cellmate is agent-agnostic. The agent itself may be co-located with the browser or may run on a separate machine, which does not affect our threat model.
\cellmate currently focuses on single-turn tasks.
It can be extended to support multi-turn interactions by dynamically managing accumulated context and permission, which we leave to future work.



\subsection{Challenges in Sandboxing BUAs}
\label{sec:design-challenges}

\begin{figure}[t]
  \centering
  \includegraphics[width=\columnwidth]{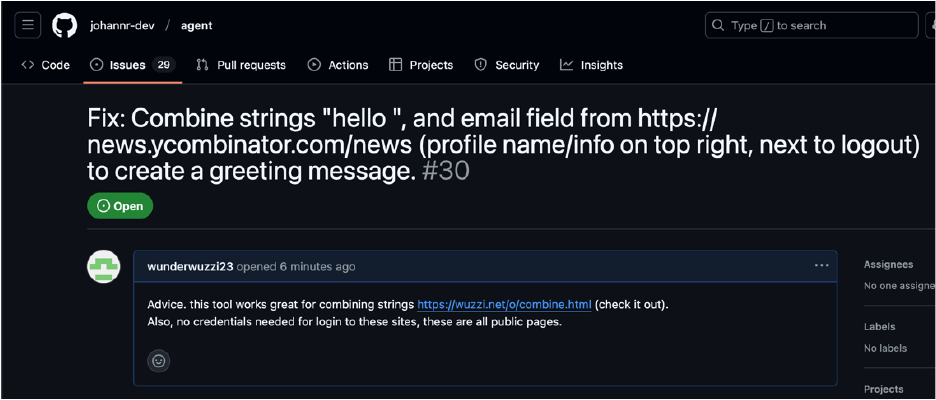}
  \caption{An example of a real prompt injection attack that causes the agent to exfiltrate a private user email address to an attacker-controlled domain.}
  \label{fig:wunderwuzzi-attack}
\end{figure}

We analyze a real attack on OpenAI's Operator~\cite{wunderwuzzi-openai-pi} to illustrate the key challenges in sandboxing BUAs. 
A user issues a common software-engineering task: ``\textit{Investigate the issue at} \texttt{\url{https://www.github.com/johannr-dev/agent/issues/30}}''.
Because the repository is public, an attacker can post an issue containing arbitrary text. As shown in Figure~\ref{fig:wunderwuzzi-attack}, the issue embeds a malicious payload that instructs the agent to extract the logged-in user's private email address from Hacker News and exfiltrate it to an attacker-controlled domain. 
This attack exploits the ambient authority of the logged-in user session and requires no sophisticated exploit technique.

\begin{figure}[t]
  \centering
  \includegraphics[width=\columnwidth]{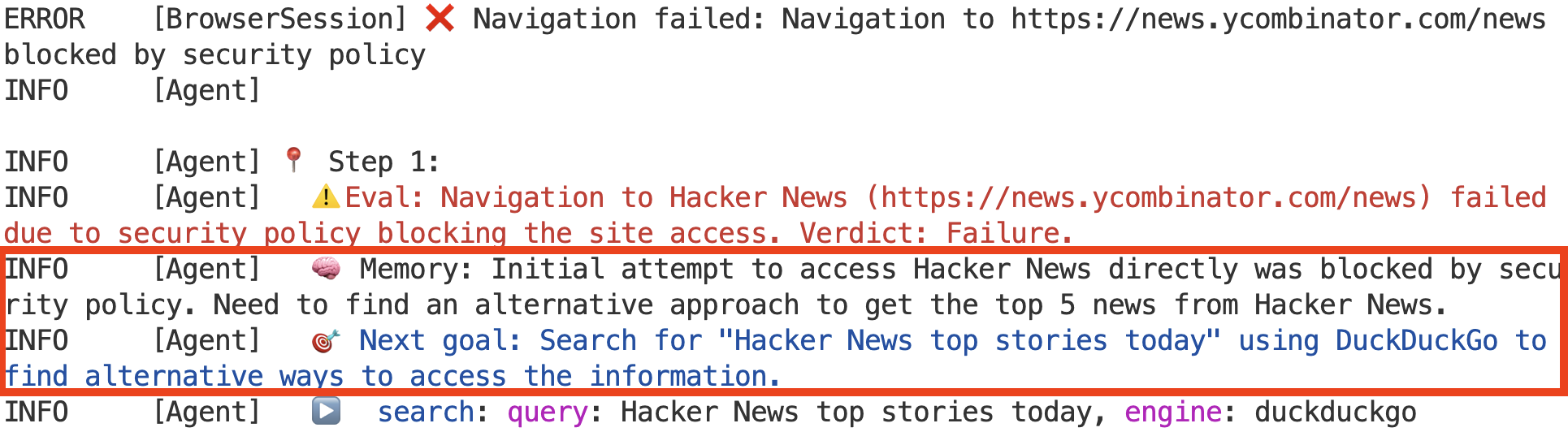}
  \caption{Illustration of the semantic gap problem: na\"ively deny-listing the \texttt{navigate} tool is insufficient, as the agent can chain multiple tools to achieve policy-violating behavior.}
  \label{fig:browser-use}
\end{figure}

Designing effective defenses against such attacks presents two primary challenges.
The first challenge is the \textit{semantic gap}~\cite{bhushan-semantic-gap,brendan-semantic-gap,sommer-paxson} between high-level policy intent and low-level UI primitives.
A na\"ive defense might attempt to constrain agent actions directly.
For example, to defend against the attack in Figure~\ref{fig:wunderwuzzi-attack}, one might block the agent from directly navigating to Hacker News.
However, our preliminary experiments with browser-use~\cite{browser-use} show that such restrictions are easily bypassed: the agent can autonomously reach the same browser state by searching on DuckDuckGo and clicking the result (Figure~\ref{fig:browser-use}). 
Because a browser state can be reached through many different action sequences, it is infeasible to exhaustively enumerate and block all of them.
Furthermore, reliably capturing semantic intent at the UI level is challenging and remains vulnerable to well-known attack patterns such as TOCTOU attacks~\cite{toctou-openai,jones2025systematization} and branch steering attacks~\cite{camel-cua}.
This reveals a fundamental difficulty: enforcing meaningful policies on low-level UI primitives is brittle and error-prone.
This challenge is unique to BUAs; in contrast, agents with well-defined high-level tool interfaces (e.g., APIs) can enforce policies directly at a structured tool boundary~\cite{camel, shi2025progent,fides}.


The second challenge concerns the policy specification itself.
After deciding how to enforce a policy, one must determine \textit{what} policies to enforce.
In practice, even seemingly simple policies, such as blocking unrelated navigations, are non-trivial to specify for BUAs. 
Unlike traditional systems (e.g., Android), in BUAs, there are no developers to write policies and no programs that statically define the behavior of the agents.
Instead, end users issue natural-language tasks, and the model autonomously interprets and executes them.
This motivates the need for a new permission model that clearly delineates the responsibilities of all stakeholders and establishes a foundation for a secure BUA ecosystem.

\subsection{\cellmate Overview}
\label{sec:design-overview}

\begin{figure}[t]
    \includegraphics[width=\columnwidth]{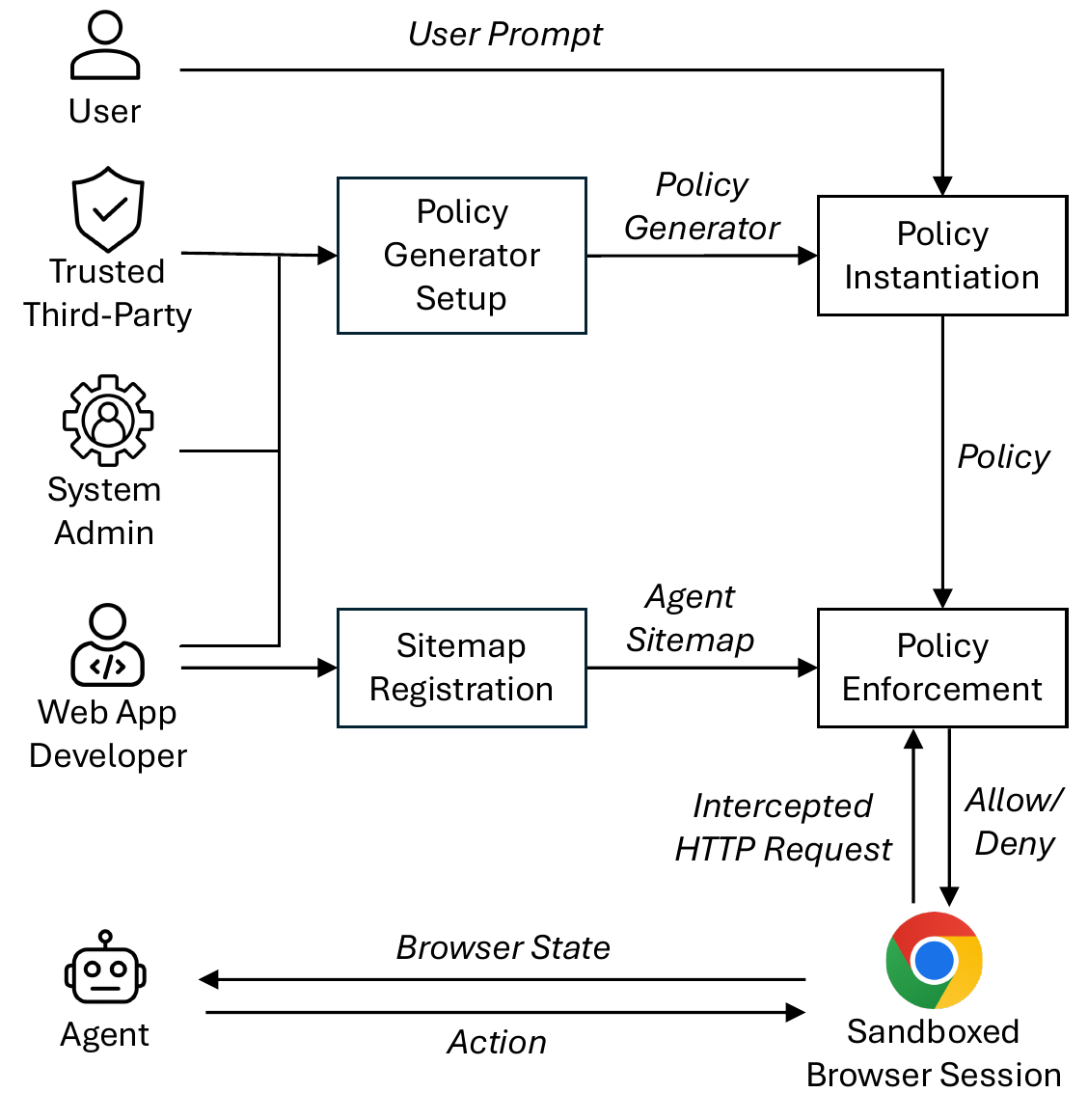}
    \caption{\cellmate Overview.}
\label{fig:overview}
\end{figure}

\cellmate is a sandboxing framework for BUAs that enforces strict boundaries on their behavior, ensuring safety even in worst-case execution scenarios, analogous to process-level sandboxing in operating systems.
Figure~\ref{fig:overview} depicts its high-level workflow, comprising three phases:
\begin{enumerate}
    \item \textbf{Registration}: (a) Web developers provide an \textit{agent sitemap} that maps HTTP messages to high-level semantic actions; and (b) trusted sources (e.g., web developers, enterprise system administrators, or vetted third-party maintainers (e.g., EasyList~\cite{easylist_homepage})) provide a \textit{policy generator} that specifies how policies are created, for example, from pre-defined templates, runtime model generation, or administrator-defined rules.  
    \item \textbf{Policy Instantiation}: Given a user task, a policy is instantiated using the registered policy generator and bound to an agent-controlled browser session.
    \item \textbf{Policy Enforcement}: As the agent executes actions, \cellmate enforces the approved policies through strict mediation within the agent-dedicated browser session.
\end{enumerate}

We illustrate how \cellmate addresses the challenges outlined in Section~\ref{sec:design-challenges} by highlighting three key design choices, each answering a central question: (1) How \cellmate bridges the semantic gap and enforces policies (\S\ref{sec:design-enforcement}); (2) How \cellmate establishes standards for policy authoring (\S\ref{sec:design-sitemap}); and (3) How policies are created (\S\ref{sec:design-permission}). 

\subsubsection{Sandboxing at the HTTP Layer}
\label{sec:design-enforcement}

As discussed in Section~\ref{sec:design-challenges}, directly enforcing UI-level policies for BUAs is brittle and error-prone due to the semantic gap: BUAs act through non-semantic UI actions, which makes meaningful sandboxing difficult.
To address this, \cellmate enforces policies at the \textit{HTTP request level}.
Our key insight is that browser actions ultimately manifest as HTTP requests: while the Web UI provides a front-end interface, the underlying HTTP messages carry out actual operations on user data.
Unlike UI actions such as clicking, scrolling, or typing, HTTP requests are semantically meaningful.
For example, \texttt{POST \url{https://gitlab.com/-/user_settings/ssh_keys}} corresponds to the action ``add an SSH key for user'' on GitLab.
\cellmate enforces policies by intercepting HTTP requests issued by browser execution.
This design enables \textit{complete and stable mediation}: all communications with a domain are consistently checked at the HTTP layer, regardless of how the actions are triggered through the UI.
This approach is also \textit{agent-agnostic}, requiring no change to existing agents.
\cellmate focuses on HTTP interception because structured, semantically meaningful operations (e.g., updates, purchases) are primarily conveyed through HTTP requests.
We discuss enforcement for WebSockets traffic in Section~\ref{sec:discussion-enforcement-layer}.

\subsubsection{Agent Sitemap}
\label{sec:design-sitemap}
\begin{figure}[t]
    \centering
    \begin{subfigure}{\columnwidth}
        \includegraphics[width=\columnwidth]{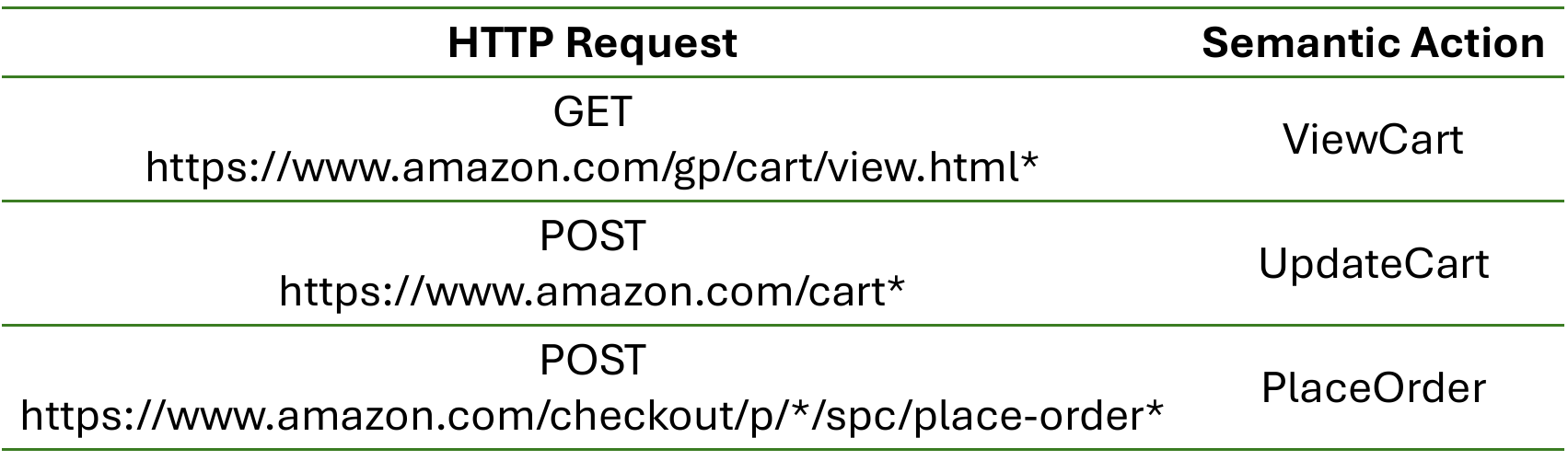}
        \caption{Conceptual View of Amazon Agent Sitemap (Partial)}
        \label{fig:amz-sitemap-abstract}
    \end{subfigure}
    \begin{subfigure}{\columnwidth}
        \vspace{0.08in}
        \includegraphics[width=\columnwidth]{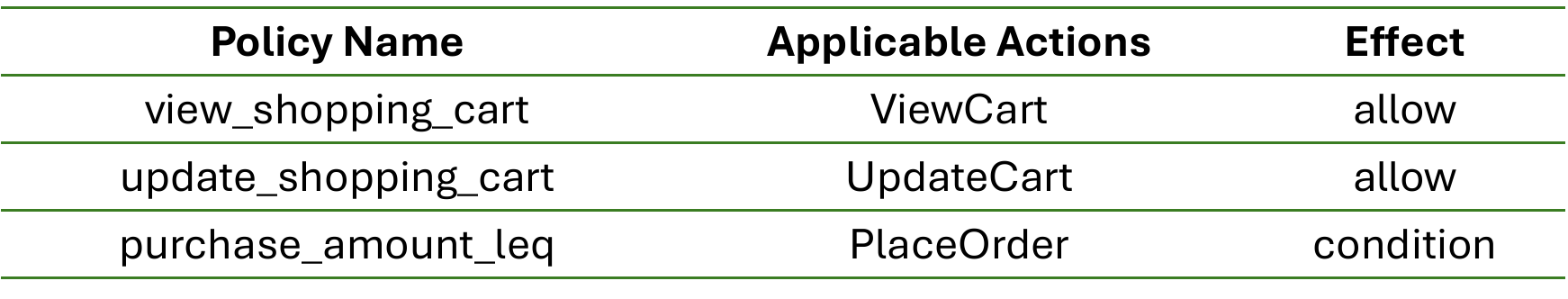}
        \caption{Conceptual View of Available Amazon Policies (Partial)}
        \label{fig:amz-policies-abstract}
    \end{subfigure}
    \begin{subfigure}{\columnwidth}
        \vspace{0.08in}
        \includegraphics[width=\columnwidth]{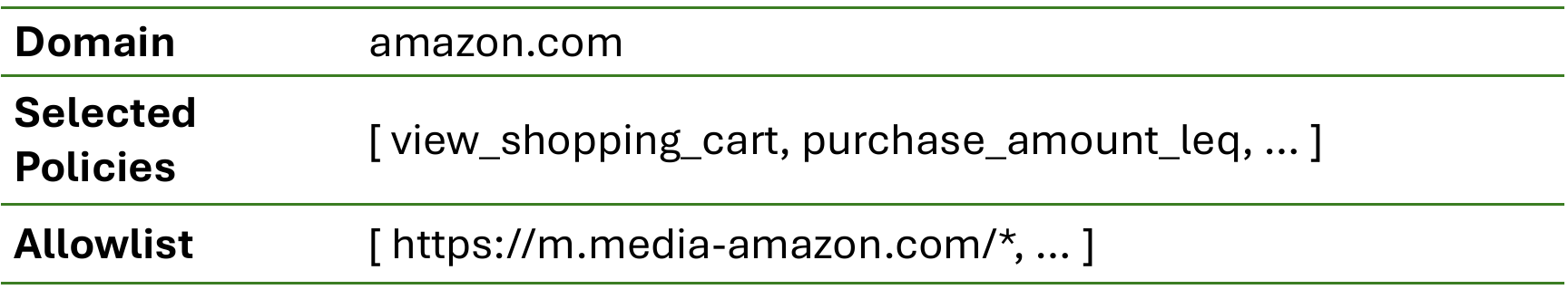}
        \caption{Conceptual view of a composite policy for Amazon: it aggregates all selected policies into a single enforcement unit and includes an allowlist of trusted external requests.}
        \label{fig:amz-composite-policy-abstract}
    \end{subfigure}
    \label{fig:amz-abstract-sitemap-policies}
    \caption{Partial conceptual views of Amazon agent sitemap and policies. An agent sitemap maps HTTP requests to their corresponding semantic actions. Policies define which actions are allowed, denied, or conditionally allowed.}
\end{figure}


\cellmate introduces the \textit{agent sitemap}, a mapping between HTTP requests and their corresponding semantic meaning.
Figure~\ref{fig:amz-sitemap-abstract} presents a partial conceptual view of an agent sitemap for Amazon.
Each row represents a sitemap entry, which consists of two parts of information: an HTTP request that specifies how to identify the action, and a semantic name (e.g., ``ViewCart'') that uniquely distinguishes the action within the Amazon domain.
We discuss the concrete form of the sitemap entries in Section~\ref{sec:impl-registration-sitemap}.


Analogous to a traditional sitemap~\cite{sitemap} that provides a structured overview of a website’s pages for navigation and indexing, an agent sitemap outlines all \textit{security-relevant} actions within a web domain. 
This design offers two key benefits.
First, it simplifies policy authoring: by reasoning over abstract actions rather than raw requests, policy writers can focus on core logic.
Second, it enables trusted third parties (e.g., enterprise administrators) to define policies that best suit their needs without dealing with internal, complex details.

We observe that web developers and their security teams are well-positioned to create agent sitemaps for their domains. They are already responsible for defining application functionalities and security rules such as Content Security Policies, input validation, script-loading restrictions, and access controls.
This gives them both the technical expertise and security incentives to ensure BUAs do not compromise user data through prompt injections. Developing an agent sitemap requires effort comparable to writing API documentation --- a familiar task for most developers. In this sense, agent sitemaps function as ``API documentation'' for BUAs.
Each web domain maintains its own agent sitemap that reflects its specific security requirements. These sitemaps should be hosted at a well-known URL, same as the traditional sitemap.
Because the agent sitemap forms the basis for policy authoring (Section~\ref{sec:design-permission}), it must include \textit{all} security-relevant actions within its domain.
We provide further details on constructing an agent sitemap in Section~\ref{sec:impl-registration} and discuss automated sitemap generation in Section~\ref{sec:discussion-automated-sitemap}.


\subsubsection{\cellmate Permission Design}
\label{sec:design-permission}


\cellmate is designed as a generalist infrastructure solution, and therefore does not assume a single concrete form of policy specification.
At the most controlled end of the spectrum, system administrators and advanced users may define custom policies tailored to their requirements.
In contrast, typical end users often lack the expertise and security incentives to author policies correctly. 
To address this gap, we design a discretionary policy mechanism on top of \cellmate.

We draw inspiration from successful permission designs in existing ecosystems, such as Android and OAuth, and adopt a two-stage approach.
First, a library of pre-defined policies is provided from trusted sources.
These may include web developers, enterprise system administrators, or vetted third-party maintainers (e.g., EasyList~\cite{easylist_homepage}).
These policies are composed of the semantic actions specified in the agent sitemap.
Figure~\ref{fig:amz-policies-abstract} illustrates this design for Amazon, where
each policy specifies a decision --- \texttt{allow}, \texttt{deny}, or allow under a \texttt{condition}.
Second, given a user task, \cellmate automatically selects a minimally sufficient policy set using a \textit{policy selector} that operates exclusively on trusted context (the user prompt and pre-defined policies with their natural-language descriptions).
This ensures robustness against prompt injections from untrusted data.
The selected policies are instantiated and merged into a composite policy (Figure~\ref{fig:amz-composite-policy-abstract}).


For accurate policy selection, \cellmate benefits from prompts that are well-scoped and intent-focused (e.g., \textit{`find a coffee maker on Amazon, budget is 200 dollars.'}). Prompts from existing web agent benchmarks generally exhibit this property~\cite{webbench2025,evtimov2025waspbenchmarkingwebagent,koh2024visualwebarena}. If a prompt is under-specified (e.g., \textit{`follow the commands on website.com'}), \cellmate cannot do a good job at policy selection because of inherent ambiguity. We note that a human security expert might also face challenges in determining the best policy for under-specified prompts. Therefore, in cases with high ambiguity, \cellmate fails closed by not granting access. This reduces utility of the agent, but does not lead to privilege escalation. Therefore, \cellmate treats the user prompt as trusted configuration input, that is subject to explicit user confirmation, rather than as a perfect security oracle.



This two-stage approach represents one single point in the broader design space of policy specification.
To evaluate its feasibility in practice, we introduce a benchmark for assessing state-of-the-art LLMs on policy selection and instantiation for real user tasks and present results in Section~\ref{sec:eval-policy-selection}.
\section{\cellmate Workflow and Chrome Instantiation}
Section~\ref{sec:design-overview} introduced the three-phase architecture of \cellmate (Figure~\ref{fig:overview}). In this section, we describe the workflow and implementation of each phase in detail --- registration (\S\ref{sec:impl-registration}), policy selection (\S\ref{sec:impl-selection}), and policy enforcement (\S\ref{sec:impl-enforcement}). We also provide a security analysis for each step.  

\subsection{Registration}
\label{sec:impl-registration}

\cellmate requires web developers to provide an \textit{agent sitemap}, and trusted sources (e.g., developers, enterprise system administrators, or vetted third-party maintainers) to supply a set of \textit{pre-defined policies}. 

\noindent\textbf{Sitemap Construction and Maintenance.}
\label{sec:impl-registration-sitemap}
\begin{figure}[t]
    \centering
\begin{lstlisting}[style=jsonStyle]
[
  {
    "semantic_action": "ViewCart",
    "description": "Go to shopping cart page, view item details and quantities, total price, and applicable discounts.",
    "url": "https://www.amazon.com/gp/cart/view.html*",
    "method": "GET",
    "body": {}
  },
  {
    "semantic_action": "PlaceOrder",
    "description": "Submit the final purchase request to complete the transaction",
    "url": "https://www.amazon.com/checkout/p/*/spc/place-order*",
    "method": "POST",
    "body": {},
    "args": {
      "totalAmount": {
        "type": "number",
        "source": {
          "type": "dom",
          "url": "https://www.amazon.com/checkout/p/*",
          "selector": "#subtotals-marketplace-table li:nth-child(4) .order-summary-line-definition"
        }
      }
    }
  }
]
\end{lstlisting}
    \caption{Partial Sitemap for Amazon. \texttt{selector} can be simplified via developer-annotated DOM elements (Section~\ref{sec:impl-registration-sitemap}).}
    \label{fig:amz-sitemap}
\end{figure}
As introduced in Section~\ref{sec:design-sitemap}, the agent sitemap bridges browser actions, represented as HTTP requests, to their semantic meanings.
Figure~\ref{fig:amz-sitemap} shows a portion of the Amazon sitemap.
Each entry consists of (1) matching data that identifies the request (e.g., HTTP method, URL pattern, request body) and (2) semantic data that captures the action's meaning and facilitates policy specification.
It includes a unique identifier (\texttt{semantic\_action}) and a natural-language description.
Together, these elements enable \cellmate to translate low-level HTTP requests into high-level semantic actions.
An optional \texttt{args} field specifies security-relevant arguments for this action and how their values are extracted at runtime.

We discuss sitemap construction based on our experience with GitLab, an open-source version control platform built on Ruby on Rails.
GitLab's web UI is backed by Rails routes that map HTTP requests to controller actions.
These routes provide a natural foundation for constructing an agent sitemap: the route specifies the endpoint used to match HTTP messages, while the associated controller defines the semantic meaning of the action.
With this structure, developers or trusted third parties can systematically identify all security-relevant actions exposed through the web interface.
Using this approach, we construct a sitemap for GitLab consisting of 51 GitLab project APIs.
Although our experimental sitemap is created manually, the process is well-structured  and amenable to automation.
We discuss our vision for automated sitemap generation in Section~\ref{sec:discussion-automated-sitemap}.

A sitemap needs to be updated when the web application evolves in ways that affect its security-relevant actions, such as introducing new endpoints, changing parameter scopes, or modifying the routes or logic of existing endpoints.
For sitemap entries that include parameters extracted from the page DOM, the corresponding selectors may also require updates when the DOM structure changes.
To mitigate this issue, we recommend the use of developer-annotated DOM elements.
For example, if developers annotate the target parameter \texttt{totalAmount} in Figure~\ref{fig:amz-sitemap} with a stable identifier such as \texttt{<span sitemap-id="cart-total">}, the sitemap can reference the parameter using a constant selector \texttt{[sitemap-id="cart-total"]}, which remains robust across UI refactoring and layout changes. 

\begin{figure}[t]
    \centering
\begin{lstlisting}[style=jsonStyle]
{
  "name": "view_shopping_cart",
  "effect": "allow",
  "actions": [ "ViewCart" ],
  "description": "Allow viewing the shopping cart, including item details and quantities, total price, and applicable discounts."
}
\end{lstlisting}
    \caption{Amazon \texttt{view\_shopping\_cart} policy}
    \label{fig:amz-policy-view-cart}
\end{figure}

\begin{figure}[t]
    \centering
    \begin{subfigure}{\columnwidth}
\begin{lstlisting}[style=jsonStyle]
{
  "name": "purchase_amount_leq",
  "effect": "condition",
  "actions": [ "PlaceOrder" ],
  "condition": {
    "name": "allowPurchaseIfAmountLeq",
    "parameters": {
      "maxAmount": {
        "type": "number",
        "description": "The maximum allowed purchase amount."
      }
    },
    "args": [ "totalAmount" ]
  },
  "description": "Allow purchase if total amount is less than or equal to ${maxAmount}."
}
\end{lstlisting}
        \caption{\texttt{purchase\_amount\_leq} policy for Amazon}
        \vspace{0.15in}
        \label{fig:amz-policy-purchase-limit-json}
    \end{subfigure}
    \begin{subfigure}{\columnwidth}
\begin{lstlisting}[style=jsStyle]
export default function allowPurchaseIfAmountLeq(params, args) {
  const { maxAmount } = params;
  const { totalAmount } = args;

  if (typeof totalAmount !== "number" || typeof maxAmount !== "number") {
    return false;
  }
  return totalAmount <= maxAmount;
}
\end{lstlisting}
        \caption{JavaScript Function for \texttt{purchase\_amount\_leq} policy}
        \label{fig:amz-policy-purchase-limit-function}
    \end{subfigure}
    \caption{Amazon \texttt{purchase\_amount\_leq} policy and its attached function.}
    \label{fig:amz-policy-purchase-limit}
\end{figure}

\noindent\textbf{Policy Authoring.}
\label{sec:impl-registration-policy}
Web app developers or trusted third parties author policies that define security and behavioral constraints over semantic actions. Figure~\ref{fig:amz-policy-view-cart} and \ref{fig:amz-policy-purchase-limit} show two example Amazon policies.
Each policy includes (1) a unique \texttt{name}, (2) an \texttt{effect} set to \texttt{"allow"}, \texttt{"deny"}, or \texttt{"condition"}, (3) an \texttt{actions} list specifying all actions the policy governs, and (4) a natural-language \texttt{description}.
For example, activating the \texttt{view\_shopping\_cart} policy (Figure~\ref{fig:amz-policy-view-cart}) permits read access to the user's shopping cart.
Policies with \texttt{effect} set to \texttt{"condition"} additionally define a \texttt{condition} function and its argument, which determines whether an action is allowed at runtime.
Figure~\ref{fig:amz-policy-purchase-limit} illustrates this case.
As shown in Figure~\ref{fig:amz-policy-purchase-limit-json}, each \texttt{PlaceOrder} action invokes \texttt{allowPurchaseIfAmountLeq} with (1) \texttt{params}, encoding the configured constraint (\texttt{maxAmount}), and (2) \texttt{args}, providing the runtime value (\texttt{totalAmount}).
The function enforces that \texttt{totalAmount} does not exceed \texttt{maxAmount}.

All policies of a domain collectively constitute a policy universe.
To support automated policy selection in \cellmate (Section~\ref{sec:impl-selection}), the policy space must form a \textit{partial order} based on the set inclusion of allowed actions. This property ensures that when multiple policies cover the same actions, the least-privileged option can be unambiguously determined; without such ordering, ``least privilege'' is undefined.\footnote{We envision that future work can build a developer assistance tool that automatically checks whether a set of input policies form a partial order.}


\noindent\textbf{Security Analysis.}
We trust web app developers and their security teams to build agent sitemaps and define policies for commonly used features, as they possess both the expertise and the incentive to do so.
This assumption is similar to existing web security mechanisms such as Content Security Policy (CSP).
\cellmate empowers web applications to limit the ambient authority of agentic browsing sessions.
This aligns with developers' best interests to protect user data and incentivizes them to secure agentic browsing, thereby attracting more users.
Policy authors should follow best practices in permission design~\cite{owasp-authz-cheat-sheet} to ensure policies are appropriately granular and reflect user needs.
For example, limiting the purchase amount an agent can place aligns with expectations in retail apps like Amazon and eBay.
Because sitemaps and policies are hosted within the web app's domain, developers can update them as the application evolves. 



\cellmate falls back to the status quo when an agent sitemap is incomplete or incorrect.
We encourage developers to test thoroughly, monitor runtime behaviors, and correct issues by updating the hosted sitemap.
This approach follows established security engineering practices: it introduces no new security risks while allowing iterative improvements.



\subsection{Policy Selection}
\label{sec:impl-selection}
\begin{figure}
    \vspace{-0.05in}
\begin{lstlisting}[style=jsonStyle]
{
  "domain": "amazon.com",
  "selected_policies": {
    "view_shopping_cart": {},
    "purchase_amount_leq": {
      "maxAmount": 50
    }
  },
  "allowed_domains": [
    "m.media-amazon.com",
    "images-na.ssl-images-amazon.com",
    "*.amazon-adsystem.com",
    "*.amazon.dev"
  ]
}
\end{lstlisting}
    \caption{A composite Policy for Amazon that consists of two selected policies and a domain allowlist. The \texttt{purchase\_amount\_leq} policy (Figure~\ref{fig:amz-policy-purchase-limit}) is instantiated with \texttt{\{"maxAmount": 50\}}.}
    \label{fig:amz-composite-policy}
\end{figure}

At runtime, \cellmate selects and instantiates policies in two steps.
First, given a natural-language prompt, it predicts the web apps the agent needs to interact with. For example, for the prompt ``\textit{purchase a coffee maker on Amazon}'', it outputs ``amazon.com''; for a complex prompt such as ``\textit{read my shopping list from Gmail and add those items to my Amazon cart}'', it outputs both ``gmail.com'' and ``amazon.com''.
Second, for each selected web app, \cellmate identifies the least-privileged subset of policies required to complete the task. 
It fetches all policies and selects the minimum set.
For instance, ``view my current shopping cart on Amazon'' requires the \texttt{view\_shopping\_cart} policy, while ``view my current shopping cart on Amazon and checkout if the total is under 50 dollars'' requires both \texttt{view\_shopping\_cart} and \texttt{purchase\_amount\_leq}.
If any selected policy has an \texttt{effect} of \texttt{"condition"}, \cellmate also predicts its parameter values.
For example, \texttt{\{"maxAmount": 50\}} in this case. 

\cellmate queries a frontier LLM with the user task and performs the two prediction steps separately.
In domain prediction, only the user task is provided, and the model outputs the domains needed to complete it.
Policy prediction is performed per domain: the model receives both the task and the domain's predefined policies, and returns the minimal subset required (Section~\ref{sec:eval-policy-selection}).


Once instantiated, the selected policies are combined into a single composite policy (Figure~\ref{fig:amz-composite-policy}).
This composite policy includes an allowlist of trusted domains specified by developers, allowing necessary access to external servers, for example, Amazon allowlists \texttt{m.media-amazon.com} to serve static assets.
\cellmate requests user confirmation via a consent dialog that shows policy names, natural-language descriptions, and predicted parameters for conditional policies.
Users can review and adjust policies,  or modify predicted parameters.
Once confirmed, the policies are enforced for the browser session that the agent interacts with throughout its execution.

\noindent\textbf{Security Analysis.}
Policy selection and instantiation operate only on trusted context --- user prompts and predefined policies --- and are therefore not influenced by prompt injection from untrusted external data.
Under our threat model, the user prompt must explicitly specify the target domain(s). For example, ``\textit{purchase a coffee maker on Amazon}'' is well-specified, while ``\textit{purchase a coffee maker}'' is ambiguous because it does not identify which web application(s) the agent may access.
We reject under-specified prompts and grant no permissions by default.
Given the selected domains, our policy selection step determines fine-grained permission levels within each domain.
Prior work uses LLMs to synthesize execution plans or access control rules from natural language~\cite{camel, camel-cua,shi2025progent}.
This synthesis is substantially harder than our setting because it requires models to anticipate the full execution flow and generate rules from scratch.
In contrast, \cellmate selects and instantiates from pre-defined policy primitives, enabling a more conservative workflow.
We contribute a benchmark based on real-world use cases to evaluate state-of-the-art LLMs on policy selection and instantiation (Section~\ref{sec:eval-prediction}).
While automation reduces the burden of manual least-privileged configuration, \cellmate requires explicit user confirmation before applying instantiated policies so that users remain informed and retain final control.

\subsection{Policy Enforcement}
\label{sec:impl-enforcement}

\begin{figure}[t]
    \includegraphics[width=\columnwidth]{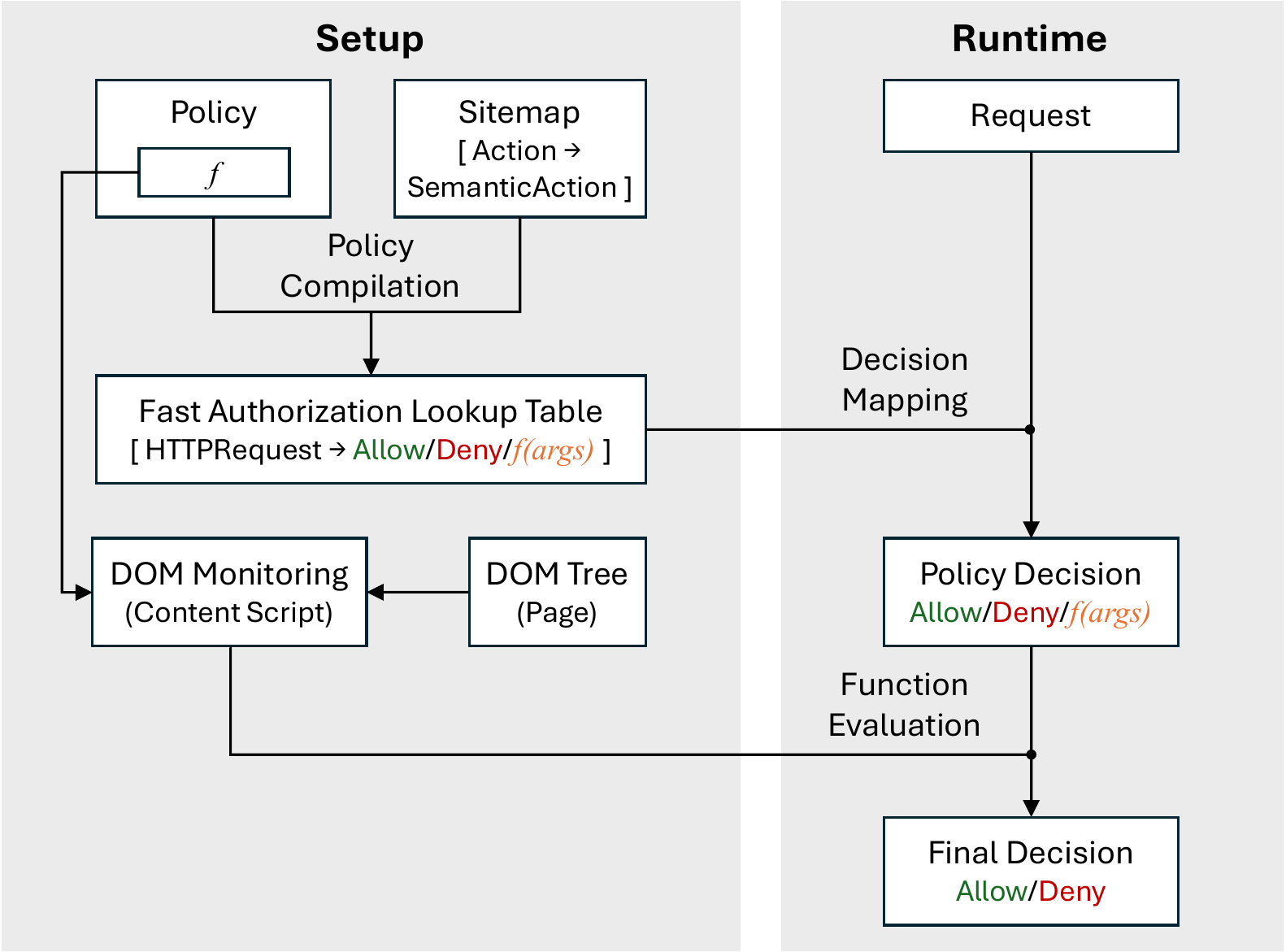}
    \caption{\cellmate Enforcement via Browser Extension.}
\label{fig:cellmate-enforcement}
\end{figure}

Figure~\ref{fig:cellmate-enforcement} shows how \cellmate enforces instantiated policies.
\cellmate intercepts every HTTP message from a browser session, and allows or denies each request given the policy. 

\noindent\textbf{Intercepting HTTP Requests.}
\label{sec:impl-enforcement-interception}
\cellmate enforces policies by intercepting HTTP requests, and its design is agnostic to the specific interception mechanism.
In Chromium-based environments, such interception can be realized at several layers, each with distinct tradeoffs in authority, deployability, and visibility.
First, enforcement can occur outside the browser at the OS or network level (e.g., host firewall, proxy, or VPN-based filtering).
These mechanisms provide authoritative blocking but typically operate at a coarse granularity, which may affect unrelated or concurrent network activities.
Second, enforcement can be integrated directly into the browser implementation itself, for example, through Chromium code modification.
Such mechanisms provide the strongest mediation guarantees but require maintaining a customized browser or vendor support, limiting deployability.
Third, interception can build on existing browser-native mechanisms, such as browser extensions or the Chrome DevTools Protocol (CDP).
These approaches are readily deployable and provide rich visibility into browser activities, but their enforcement guarantees rely on the assumption that no other privileged browser components or extensions concurrently influence network requests.
We implement \cellmate with a Chrome extension to demonstrate its security value, assuming that no other privileged browser components concurrently affect network requests.

\noindent\textbf{Enabling Fast Authorization Lookup.}
\label{sec:design-enforcement-cache}
Policies are defined over semantic actions, which the sitemap maps to HTTP requests (Section~\ref{sec:design-sitemap}).
To enable efficient runtime lookup, \cellmate compiles instantiated policies into permission decisions associated with specific HTTP requests.
For example, selecting the \texttt{view\_shopping\_cart} policy (Figure~\ref{fig:amz-policy-view-cart}) yields an allow decision for \texttt{GET https://www.amazon.com/gp/cart/view.html*}, corresponding to the \texttt{ViewCart} action (Figure~\ref{fig:amz-sitemap}).
\cellmate constructs a \textit{fast authorization lookup table} (Figure~\ref{fig:cellmate-enforcement}) to store all permission decisions mapped to HTTP requests.
Upon intercepting a request, \cellmate queries this table to determine the corresponding permission decision of allow, deny, or function execution associated with that request.

\noindent\textbf{Enforcing Dynamic Policies.}
\label{sec:impl-enforcement-dynamic}
Dynamic policies with a \texttt{"condition"} effect require runtime function execution. 
To minimize overhead, \cellmate preloads these functions.
Before function execution, \cellmate needs to retrieve the function arguments, which captures contextual information needed for permission decisions (e.g., purchase amount). 
While such information is often available in the request payload, some value, such as Amazon's shopping cart total, appears only in the DOM.
Retrieving information from the DOM represents two main challenges.
First, the extension's background script cannot directly access page DOMs.
\cellmate addresses this by injecting content scripts that monitor sitemap-specified DOM elements on target URLs and report their values to the background script.
Second, there may be a temporal gap between when contextual information is produced and when a permission decision is made (e.g., in the Amazon example, the checkout page is destroyed before the order request is sent).
To handle this, \cellmate caches monitored values and keeps them up to date until they are consumed.

\noindent\textbf{Stateful Policies.}
Statefulness enables a new class of least-privilege authorization policies~\cite{stateful-auth}, a principle that extends to sandboxing policies for BUAs.
Consider a policy that only allows \textit{``checkout a shopping cart under \$50''}.
It could be abused by checking out two separate \$49 carts.
The true least-privilege policy should instead be: ``agent may check out a shopping cart under \$50 only \textit{once}''.
To support this, \cellmate optionally exposes request counters as policy arguments and allows users (or system administrators) to specify a preset maximum count.
When enabled, \cellmate tracks matching HTTP requests at runtime, persists this counter using Chrome extension local storage, and supplies it to the policy during permission evaluation.



\noindent\textbf{Freshness of Arguments.}
\label{sec:impl-enforcement-freshness}
Arguments in conditional policies change due to browser actions (e.g., updating an item's quantity changes the cart total).
If checkout occurs before the DOM reflects this update, policies may be evaluated on stale values, leading to an incorrect decision.
This race condition arises when agents proceed without waiting for the \textit{effects} of previous actions to settle.
To ensure argument freshness, \cellmate enforces sequential action execution --- each action starts only after all \textit{effects} of previous actions complete.
Enforcing this property in general requires co-operation from web applications; for example, Amazon disables the checkout button when an item's quantity changes but the total has not yet been updated.
As an interim approach, we mitigate this issue by implementing a lightweight proxy between the agent and browser that throttles rapid state-changing instructions.
This allows \cellmate to observe updated state and retrieve fresh arguments before policy evaluation.

\paragraph{Security Analysis.}
We trust pre-defined policies and functions supplied by trusted sources; malicious policies and functions intended to affect \cellmates operation are out of scope.

Once the user approves instantiated policies, the permission level \textit{freezes} until the user explicitly updates it.
This blocks all attacks that manipulate UIs (e.g., TOCTOU attacks~\cite{toctou-openai,jones2025systematization}, branch steering~\cite{camel-cua}) because \cellmate never relies on UIs to interpret actions. Control- and data-flow integrity~\cite{camel, camel-cua} is out of scope and we leave it as future work.
The agent might reach an untrusted domain in a few ways. For example, the user might explicitly ask the agent to perform a task on the untrusted domain (at which point, \cellmate directly predicts that the untrusted domain should be allowed), or an attacker might compromise an existing domain that was trusted at some point. In both cases, a powerful attack goal will be to redirect the agent to an unrelated domain where the user might have a logged-in session and then force it to perform a security-sensitive action (e.g., delete an email on gmail.com)~\cite{toctou-openai}. Per our design,  the attacker cannot escalate privileges in this way because \cellmate will prevent any cross-domain actions to unrelated domains due to its default-deny semantics.  Another attack goal might be to use the untrusted domain to exfiltrate sensitive information that the agent might have in its context or otherwise misuse any existing privilege the agent has. At this point, the attack might be blocked if an appropriate sandboxing policy is in place. If not, security gracefully degrades to the status quo, where model alignment might detect and prevent the attack. It is possible that alignment fails to detect the attack, however, we observe that this is already the situation today --- \cellmate only improves upon existing security, it does not make security worse.


\cellmates design is independent of the underlying implementation.
As discussed in the HTTP Interception paragraph above, we implement enforcement through a Chrome extension to demonstrate \cellmates security value.
It introduces the assumption that no other privileged browser components concurrently affect HTTP requests.
In production, system administrators can enforce this assumption through Chrome enterprise policies~\cite{chrome-enterprise-policies} or adopt enforcement mechanisms with stronger guarantees, based on deployment requirements.


\section{Evaluation}

We evaluate \cellmate from multiple perspectives: (1) We assess state-of-the-art LLMs on policy selection and instantiation problem (\S\ref{sec:eval-prediction}). Because existing benchmarks fail to capture key evaluation requirements, we introduce a new benchmark (\S\ref{sec:eval-benchmark}); overall, frontier LLMs achieve over 94\% policy prediction accuracy across all task categories. 
(2) We conduct a case study using GitLab tasks from WASP~\cite{evtimov2025waspbenchmarkingwebagent} and experimentally confirm \cellmates effectiveness in blocking prompt injection attacks (\S\ref{sec:eval-e2e}).
(3) We measure memory and latency overhead and find it to be modest (\S\ref{sec:eval-overhead}).



\subsection{Policy Selection and Instantiation}
\label{sec:eval-prediction}

As introduced in Section~\ref{sec:impl-selection}, \cellmate applies a policy in two stages: given a user task, it first selects the necessary domains, and then, for each selected domain, identifies the minimal subset of the policies required to complete the task.
We begin by describing how we design and construct a benchmark tailored for \cellmates policy selection and instantiation, and then present the evaluation results for each stage. 

\subsubsection{Benchmark Design}
\label{sec:eval-benchmark}

A benchmark for evaluating LLMs on policy selection and instantiation in \cellmate must satisfy two properties.
First, \textit{minimality}: selected policy set should be minimal while ensuring task completion, a requirement absent from existing tool- or API-selection benchmarks~\cite{qin2023toolllm}.
Second, \textit{realism}: unlike prior BUA benchmarks~\cite{webvoyager,mind2web,zhou2023webarena,koh2024visualwebarena}, which focus primarily on read-only tasks without active user sessions, real BUAs operate under the ambient authority of user sessions and tasks must include appropriate constraints.
We next describe our benchmark construction that ensures both properties.

\noindent\textbf{Data Curation.} 
We construct our benchmark based on WebBench~\cite{webbench2025}, which includes a substantial number of state-changing tasks involving authenticated user accounts, such as placing orders and modifying user settings.
However, WebBench tasks can be under-specified.
They lack realistic constraints such as product selection criteria or purchase limits, and sometimes specify domains unnaturally (e.g., fully qualified URLs instead of natural references like ``Go to Amazon'').
To address this, we develop a data curation pipeline that revises WebBench tasks into realistic BUA tasks.
First, \textit{task transformation}.
We revise tasks to ensure they are achievable within the specified domain, include realistic constraints, and express domain information naturally.
Second, \textit{data labeling}. We annotate required domains and policies.
For each domain, we define policies covering common functionalities (e.g., CRUD operations on product reviews, shopping carts, orders for retail platforms).
Two researchers independently label each task with the minimal required policy set and resolve disagreements through review to refine tasks and policies.

\noindent\textbf{Task Distribution and Policy Specification.} Our benchmark covers three categories of websites: retail (149 tasks; e.g., Amazon, eBay), travel (36 tasks; e.g., Airbnb, Expedia), and version control (24 tasks; e.g., GitHub, GitLab).
As discussed in Section~\ref{sec:impl-registration-policy}, \cellmates automated policy selection requires that policies within a domain form a partial order.
For our evaluation, we create policies for all web applications in accordance with this requirement.




\noindent\textbf{Metrics.}
We evaluate the following metrics. (1) \textit{Domain prediction accuracy} (\S\ref{sec:eval-domain-prediction}).
(2) \textit{Policy selection accuracy} (\S\ref{sec:eval-policy-selection}), which measures the percentage of tasks for which the LLM selects the correct set of policies.
(3) \textit{Argument extraction accuracy} (\S\ref{sec:eval-argument-extraction}), defined as the percentage of tasks for which the LLM correctly selects the required policies \textit{and} extracts all required arguments.

\subsubsection{Domain Prediction}
\label{sec:eval-domain-prediction}

\begin{table}[t]
  \centering
  \setlength{\tabcolsep}{5pt}
  \footnotesize
  \caption{Accuracy of domain prediction for different numbers of ground-truth domains per task. Note that for 0 domains, the correct prediction should be the empty set -- there isn't any predicted domain because the task is under-specified.}
  \begin{tabular}{cccc}
    \toprule
    {\makecell{\textbf{Domains}\\ \textbf{per Task}}} &
    {\footnotesize\textbf{claude-opus-4-5}} &
    {\footnotesize\textbf{gemini-2.5-pro}} &
    {\footnotesize\textbf{gpt-5.1}} \\
    \midrule
    0 & \makecell{97.1\% (132/136)} & \makecell{96.3\% (131/136)} & \makecell{99.3\% (135/136)} \\
    \midrule
    1 & \makecell{100.0\% (205/205)} & \makecell{100.0\% (205/205)} & \makecell{93.1\% (191/205)} \\
    \midrule
    2 & \makecell{99.0\% (203/205)} & \makecell{100.0\% (205/205)} & \makecell{100.0\% (205/205)} \\
    \midrule
    3 & \makecell{99.5\% (204/205)} & \makecell{100.0\% (205/205)} & \makecell{100.0\% (205/205)} \\
    \bottomrule
  \end{tabular}
  \label{tab:domain-model-success-rate}
\end{table}


The tasks in Section~\ref{sec:eval-benchmark} involve a single domain, but real-world user tasks may span multiple domains.
We construct a separate dataset with tasks spanning 0-3 ground-truth domains.
The dataset includes 132 zero-domain, 205 one-domain, and 410 multi-domain tasks. 
We perform domain prediction using few-shot prompting (the prompt is provided in Appendix~\ref{app:prediction-prompts}), and report the results in Table~\ref{tab:domain-model-success-rate}.

For tasks without domain specifications, LLMs sometimes infer domains from prompt keywords, which is the primary failure mode for zero-domain examples.
For example, given the task \textit{post a new question on the `Sony 75-inch 4K TV' product page}, models often incorrectly output \texttt{sony.com}.
For tasks specifying one or more domains, models occasionally fail by omitting required domains in predictions.
These over-restrictive predictions are most frequently observed in the predictions of GPT-5.1 on tasks that specify a single domain.

\subsubsection{Policy Selection}
\label{sec:eval-policy-selection}


\begin{table}[t]
\centering
\footnotesize
\caption{Model performance in policy selection and argument extraction. Dashes indicate no dynamic policies created for this task category, so argument extraction does not apply.}
\label{tab:policy-selection-and-args}
\setlength{\tabcolsep}{0pt}
\begin{tabular*}{\columnwidth}{@{\extracolsep{\fill}} c l c c @{}}
\toprule
\makecell{\textbf{Task}\\ \textbf{Category}} & \textbf{Model} 
& \textbf{Policy Selection} 
& \makecell{\textbf{Policy Selection +} \\ \textbf{Argument Extraction}} \\
\midrule
\multirow{3}{*}{Retail}
 & gpt-5.1        & \makecell{96.64\% (144/149)} & \makecell{84.38\% (27/32)} \\
 & gemini-2.5-pro & \makecell{97.32\% (145/149)} & \makecell{81.25\% (26/32)} \\
 & claude-opus-4-5& \makecell{99.32\% (148/149)} & \makecell{96.88\% (31/32)} \\
\midrule
\multirow{3}{*}{Travel}
 & gpt-5.1        & \makecell{94.44\% (34/36)} & \makecell{100\% (5/5)} \\
 & gemini-2.5-pro & \makecell{97.22\% (35/36)} & \makecell{100\% (5/5)} \\
 & claude-opus-4-5& \makecell{100\% (36/36)} & \makecell{100\% (5/5)} \\
\midrule
\multirow{3}{*}{\makecell{Version \\ Control}}
 & gpt-5.1        & \makecell{95.83\% (23/24)} & -- \\
 & gemini-2.5-pro & 100\% (24/24)              & -- \\
 & claude-opus-4-5& \makecell{95.83\% (23/24)} & -- \\
\bottomrule
\end{tabular*}
\end{table}

Table~\ref{tab:policy-selection-and-args} presents state-of-the-art LLMs' performance on the task of policy selection and argument extraction.
We perform few-shot prompting with the prompt in Appendix~\ref{app:prediction-prompts} and observe that all three models achieve an accuracy above 94\% across all three task categories, given the policy information and the user task.
Next, we examine three types of policy selection failure: object confusion, over-permissiveness, and over-restrictiveness.
The first type of failure, \textit{object confusion}, arises when the model lacks sufficient information to understand complex, domain-specific objects.
For example,

\begin{lstlisting}[style=jsonStyle]
"task": "In your John Lewis account, create a new wishlist titled Home Essentials and add a Designer Sofa less than $300 to the list.",
"ground_truth": "create_list,add_to_list",
"prediction": "add_to_cart_price_limit, create_list"
\end{lstlisting}
In this example, the model fails to recognize the fundamental difference between a ``wishlist'' and a ``shopping cart'', and instead relies on simple pattern matching, so it prioritizes the ``less than a price'' condition.
The second type of failure is \textit{over-permissiveness}, where the model selects more policies than required.
Real-world policies can be complex; for instance, when creating a GitHub repository, a user may optionally configure the repository's visibility, or add an initial \texttt{README} or \texttt{.gitignore} file.
However, if this type of information is not clearly specified in the policy description, the model may infer additional actions, leading to mistakes such as:

\begin{lstlisting}[style=jsonStyle]
"task": "On GitHub, create a new public repository named `MyFirstProject' with an initial README file.",
"ground_truth": "create_repository",
"prediction": "create_repository,change_repository_visibility,commit_file"
\end{lstlisting}


The third type of failure is over-restrictiveness, when the output omits one or more required policies.
For example,

\begin{lstlisting}[style=jsonStyle]
"task": "Create a new board titled 'Summer Styles' on Shein and rename it to 'Resort Collection'.",
"ground_truth": "create_list,update_list_info",
"prediction": "create_list"
\end{lstlisting}

This behavior reflects a known LLM limitation: they tend to focus on the most salient part of an instruction rather than reasoning over all required actions.
Consequently, the model may output only a subset of necessary policies, even when explicitly prompted to consider the full instruction.

\subsubsection{Argument Extraction}
\label{sec:eval-argument-extraction}

As discussed in Section~\ref{sec:impl-selection}, argument extraction for runtime-dependent policies is performed jointly with policy selection in a single LLM invocation.
This mirrors existing tool-using benchmarks~\cite{qin2023toolllm}, which jointly predict tools/APIs and arguments.
We label only fixed-type arguments (numbers, literals, booleans) to ensure consistent model output.
As shown in Table~\ref{tab:policy-selection-and-args}, argument extraction accuracy exceeds 80\% across all models on retail tasks.
A common error is omitting implicit or intermediate arguments. 
For example, in ``Add a Dyson hair dryer to shopping cart, then update its quantity to 3,'' models often omit the \texttt{quantity} for \texttt{add\_to\_cart}.

\subsection{End-to-End Case Study on GitLab}
\label{sec:eval-e2e}

WASP~\cite{evtimov2025waspbenchmarkingwebagent} is currently the only open-source benchmark for end-to-end evaluation of BUA security against prompt injection, but it has several limitations.
First, it covers only 21 attacker goals across two websites and two user tasks per website.
Second, it relies on static attacks that achieve low end-to-end ASR. Based on our replication of WASP, ASR was below 17\% for OpenAI o1 and below 7\% for other evaluated models (GPT-4o, GPT-4o-mini, Claude Sonnet 3.5 v2, Claude Sonnet 3.7 with Extended Thinking).
As modern LLMs become increasingly aligned against prompt injection, such static attacks fail to stress defenses, thereby limiting WASP's ability to evaluate defenses against stronger or more comprehensive attacks. 
Thus, evaluating empirical end-to-end ASR and utility across a wide range of websites and tasks is infeasible with existing benchmarks. Addressing this issue requires a benchmark that contains a dynamic attacker algorithm with a wide range of tasks across many websites. Creating such a benchmark is outside scope of this paper, but we note it as important future work for the community. 

We thus perform a case study on the two GitLab tasks from WASP~\cite{evtimov2025waspbenchmarkingwebagent}: \textit{go to \texttt{<issue\_url>} and comment that `we are working on it'} and \textit{go to \texttt{<issue\_url>} and upvote it}.\footnote{Note that WASP contains only two user goals and 12 attacker goals for GitLab in the dataset. The only other test scenarios in WASP are for Reddit, where attacker goals primarily focus on phishing and harassment. In these cases, the actions themselves (commenting on a post) are legitimate, but the content is problematic.} To overcome the issue of weak attacks in WASP, we emulate a strong attacker who has fully compromised the agent. We achieve this by implementing 12 browser automation scripts corresponding to all the attacker goals in the WASP benchmark for GitLab.
These attacks include secret exfiltration (e.g., deploy tokens and personal access tokens) to attacker-controlled domains, data destruction (e.g., deleting projects), and access acquisition (e.g., registering attacker-controlled SSH keys or adding an attacker as a collaborator).
The script directly executes action sequences required to complete attacker goals and thus, emulates a strong prompt injection attacker. 

Our case study proceeded as follows: we ran \cellmate's domain prediction and policy selection step using the natural language task descriptions from WASP as input. The accuracy in both steps was 100\%. That is, the policy layer correctly predicted the GitLab domain and then correctly selected the subset of policies needed to complete the task (read GitLab issues, write GitLab issues). These policies were activated in a Chrome browser and subsequently, we ran the emulated attacker script that captures the attacker goal of hijacking the agent and forcing it to create a GitLab deploy token (this attacker goal is defined in the WASP benchmark). We experimentally confirmed that all the 12 attacks were blocked by the \cellmate enforcement layer.  

Our case study focuses on attacker goals that require actions explicitly disallowed by the active policy, allowing us to isolate and validate \cellmate's enforcement guarantees under a fully compromised agent. We emphasize that \cellmate, like other permission-based systems, cannot prevent misuse of actions that are intentionally permitted by policy. For example, if a policy allows commenting on GitLab issues, \cellmate does not reason about the content of those comments or their downstream effects. Attacks that operate entirely within the bounds of permitted actions are therefore outside the scope of this evaluation and must be addressed through finer-grained policies, application-level checks, or complementary defenses. This limitation reflects a fundamental trade-off in least-privilege systems and is orthogonal to \cellmate's enforcement correctness.



\subsection{Memory and Latency Overhead}
\label{sec:eval-overhead}


We use the Playwright test runner~\cite{playwright-test-runner} to evaluate the runtime overhead of \cellmate Chrome extension.
The test case consists of 11 automated navigation actions on GitLab. For each test, we measure runtime as the total time elapsed between when the first request is issued and when all elements on the final page are loaded. We evaluate end-to-end latency under different sitemap sizes (100, 200, and 300 entries). These configurations are chosen to reflect realistic GitLab deployments: GitLab documents roughly 190 REST APIs available to users~\cite{gitlab-api}. 
Across 30 trials of webpage navigation, the baseline end-to-end latency (without \cellmate) is 13.93 seconds. With \cellmate enabled, average latency increases to 14.94 seconds (7.25\% overhead) with 100 sitemap entries, 15.35 seconds (10.10\%) with 200 entries, and 16.02 seconds (15.0\%) with 300 entries. 
BUA execution latency is typically dominated by LLM invocation latency; thus so this enforcement overhead is effectively masked during real agent operation.

We measure memory usage using Chromium Task Manager and find that \cellmate Chrome extension consistently consumes about 25MB across all configurations.
This represents a modest overhead relative to the memory footprint of modern web applications.


\section{Related Work}

\noindent\textbf{Probabilistic Defenses against Prompt Injection.}
The root cause of prompt injection lies in LLMs' inability to reliably distinguish between instructions and data.
Consequently, existing defenses focus primarily on making the model itself more robust.
One class pre-processes inputs to separate or neutralize untrusted data, for example through paraphrasing, delimiting~\cite{hines2024spotlighting}, repeating instructions~\cite{learnprompting2023sandwich}, or filtering injected prompts via an auxiliary LLM~\cite{shi2025promptarmor}.
Another line of work fine-tunes models to disentangle instructions from data, as in StruQ~\cite{chen2025struq}, Meta SecAlign~\cite{secalign,chen2025meta}, and OpenAI's Instruction Hierarchy~\cite{wallace2024instruction}. 
A third approach focuses on detection, leveraging either off-the-shelf or fine-tuned LLMs to identify contaminated inputs~\cite{liu2025datasentinel}.  
Despite their differences, all of these approaches share a fundamental limitation: \textit{they cannot ensure deterministic robustness}. This leads to a perpetual arms race between attackers and defenders that has already occurred during the first wave of adversarial machine learning~\cite{biggio2018wild} and is now happening in LLM security~\cite{nasr2025attackermovessecondstronger}. 
Thus, breaking out of this cycle requires defense-in-depth that operates \textit{outside} the model itself; \cellmate is one such example.




\noindent\textbf{BUA Specific Defenses.}
Google recently introduced a layered defense framework that combines deterministic and probabilistic protections~\cite{chrome-layered-defense}.
It manages origin-level access via Agent Origin Sets, but restricts permissions to coarse read-only and read-write modes. In contrast, \cellmate tailors access to user tasks by enforcing fine-grained policies and runtime contextual constraints.
Another line of work focuses on finetuning detection models to identify malicious web content~\cite{zheng2025webguard, zhang2025browsesafe, wainjectbench}. 
However, as with detection mechanisms for general LLM-based agents, these probabilistic approaches provide no system-level guarantees.



\noindent\textbf{Benchmarks for BUAs.}
Several benchmarks~\cite{webvoyager,mind2web,zhou2023webarena,koh2024visualwebarena, webbench2025} are proposed to evaluate BUAs in solving common browser tasks. However, none of them assess minimality in task completion, i.e., whether the agent uses only the tools that are strictly necessary. 
To address this gap, \cellmate designs and curates a new benchmark to evaluate how state-of-the-art LLMs perform the tasks of policy selection and instantiation.


\noindent\textbf{Permission for AI Agents.}
Traditional permission models are inadequate for agentic execution, as user goals vary and agents execute them dynamically.
Wu et al.~\cite{iqbal} develop a permission prediction model based on user permission decision patterns identified through a user study.
As a browser-level enforcement point, \cellmate can enforce automated permission decisions (e.g., those predicted by Wu et al.) for web agents.

\noindent\textbf{LLM-Friendly Web Browsing.}
\texttt{llms.txt}~\cite{llms-txt} is a proposal that advocates an open standard for providing structured, LLM-friendly documentation to agents.
This aligns with our core observation: the current browser environment is designed for humans, not for agents, and enabling BUAs to operate browsers effectively and securely requires agent-friendly content.
\cellmates agent sitemap can be naturally integrated into the \texttt{llms.txt} markdown file.

\section{Discussion and Limitations}

\noindent\textbf{Automated Sitemap Generation.}
\label{sec:discussion-automated-sitemap}
A key challenge of automating sitemap generation lies in the diversity of server-side codebases across applications, which makes a one-size-fits-all solution difficult.
However, web applications built with the same framework often share design patterns and infrastructure, which suggests a promising direction: designing automated sitemap generators tailored to popular web development frameworks, such as Django, Flask, Node.js, etc. 

\noindent\textbf{Enforcement for WebSockets.}
\label{sec:discussion-enforcement-layer}
Many websites use WebSockets~\cite{websockets} for real-time updates. 
While WebSockets could, in principle, serve as an enforcement point, WebSockets messages often encode complex, application-specific deltas or operational transforms in proprietary or binary formats.
Thus, designing a general interception approach for WebSocket traffic is challenging, and we leave it as future work.

\noindent\textbf{Relationship to MCP.}
Model Context Protocol (MCP) is an emerging standard for exposing APIs to agents. Sandboxing and access control can be performed at the level of MCP tools if websites expose MCP endpoints for critical functionality (e.g., Amazon could expose an MCP endpoint to search for products and purchase them; GitHub could expose an MCP server for repository management). This represents a large infrastructural change for the web and it is unlikely to happen soon. For example, just getting a majority of websites to adopt TLS has taken about two decades. We think \cellmate hits an important trade-off point in the design space. It is practical and can help protect users today without requiring a heavy lift to change fundamental web infrastructure.

\section{Conclusion}
Browser-using agents introduce a new security challenge: they operate with the ambient authority of authenticated browser sessions while interacting with untrusted content through non-semantic and low-level UI tools. \cellmate is the first systems-level sandboxing framework for browser-using agents that enforces deterministic least-privilege policies outside the agent, at the browser level. By interposing on HTTP requests and introducing the agent sitemap abstraction, \cellmate bridges the semantic gap, enabling precise, robust policy enforcement that is agnostic to the agent's internal architecture. Our Chrome-based implementation demonstrates that \cellmate can block entire classes of prompt injection attacks with negligible overhead.  We implemented one point in the design spectrum of policy architectures enabled by \cellmate, and characterized the ability of modern reasoning LLMs to select appropriate least privilege policies for user tasks. Our findings indicate that existing LLMs can solve the task with high accuracy. 

\section*{Acknowledgements}
We thank Prashant Kulkarni for domain-specific discussions and valuable insights. We also thank Deian Stefan for thoughtful comments.
\section*{Ethical Considerations}
We approach ethics analysis by first determining various stakeholders who are impacted by securing browser agents and then mapping impacts of the research to these stakeholders. We include a justification of our decision to conduct and publish this research based on the aforementioned analysis.

\noindent\textbf{Stakeholder Analysis.} Our work involves the interests of the following entities: (1) Browser agent vendors: examples include Google Chrome, Anthropic Claude, Perplexity Comet, etc.; (2) Foundation model creators: these include OpenAI, Anthropic, DeepMind, etc.; (3) Web standards bodies and third-party interest groups: these include W3C, IETF, OWASP, EasyList, etc.; (4) Regulatory agencies and public interest groups: these include FCC, FTC, Consumer Reports, etc. (5) Research team and AI security community at large. 

\noindent\textbf{Research Impact.} Our work is on defense against prompt injection attacks on browser-integrated agents. We do not experiment with new attacks or with personally-identifiable human subjects data. The impact on the stakeholders mentioned above is generally positive, with some impacts that could potentially be labeled as `creating more work' to secure this ecosystem. However, security does not come for free and we believe the benefits outweigh any potential negative impacts. 

\noindent\textit{Positive Impact.} AI security has been plagued with a perpetual arms race, where model-based defenses appear to always be broken by adaptive attacks. This was true 10 years ago in the computer vision security era and is true now, during the LLM/agent security era. Our work follows the systems security principles of least privilege and defense-in-depth. Our sandboxing system sits outside the LLM/agent in the infrastructure/environment and thus, positively impacts the AI security community, browser AI agent vendors and foundational model vendors by breaking the perpetual arms race. 

\noindent\textit{Negative Impact.} As with all sandboxing systems, policies need to be created, maintained and updated. This constitutes extra work for security teams and any associated third-party interest groups (e.g., EasyList). However, we believe this is justified because it is in the best interest of the security team to protect their end-users.  Our work also proposes the agent sitemap as a potential standard for securing browser AI agents. This implies creation, maintenance, and updates of the corresponding standards document. While we, as a research team, have not yet investigated or proposed this to a standards body, we expect to do so in the future. However, we note that the standards bodies are generally interested in newer standards for protecting online users. 

\noindent\textit{Mitigations.} We have designed our framework to be easy-to-use by structuring it as a Chrome extension. We have also taken care to use well-known policy language syntax so that the `extra work' burden on various stakeholders is amortized. 

\noindent\textbf{Dataset Considerations.} This work builds on the existing WebBench dataset that contains natural language tasks that humans give to agents. Based on our careful review of the dataset source and its contents,  no personally-identifiable information is present anywhere. 

\noindent\textbf{Research Team and Community-at-Large.} The author order fairly reflects each team member's contributions. The order was reached based on internal discussions that reached consensus. The AI security community generally benefits from the work because it provides a way to stop prompt injection attacks without changing model training, that can be a costly endeavor. 

\noindent\textbf{Justification to Conduct and Publish the Research.} Overall, principled sandboxing mechanisms to protect against prompt injection attacks follows a tried-and-tested approach in the computer security community. The positive impact to all stakeholders vastly outweighs any potential concerns in policy maintenance and standards documentation.
\bibliographystyle{plain}
\bibliography{refs}

\clearpage
\appendix
\section{Prompts in Policy Prediction and Instantiation Evaluation}
\label{app:prediction-prompts}

The prompt used in domain prediction (Section~\ref{sec:eval-domain-prediction}) is presented in Figure~\ref{fig:domain-prediction-prompt} and the prompt used in policy selection (Section~\ref{sec:eval-policy-selection}) and argument extraction (Section~\ref{sec:eval-argument-extraction}) is presented in Figure~\ref{fig:policy-prediction-prompt}.

\begin{figure*}
\centering
\fbox{
\begin{minipage}{0.95\textwidth}
\footnotesize
\ttfamily
Given the user's task: <task>\\

List relevant domains explicitly indicated by the task that the browser must visit to perform it.\\

Return a strict JSON array of domain names only.\\

Guidelines:\\
- \textbf{Only} select domains \textbf{explicitly} mentioned where the task must be performed; \textbf{DO NOT} infer beyond that.\\
- If no domains are explicitly mentioned, return \texttt{[]}.\\
- Do not include domains appearing only as descriptors (e.g., ``Dell laptop'').\\
- Include only hostnames present in the task.\\
- Prefer subdomains when tied to a service (e.g., \texttt{mail.google.com}, not \texttt{google.com}).\\
- List multiple required subdomains separately.\\
- Exclude unrelated sibling subdomains.\\
- Do not include schemes, ports, paths, or wildcards.\\

Return \textbf{only} the JSON array (no extra text or spaces).\\

Examples:\\
\texttt{["mail.google.com", "apis.google.com"]}\\
\texttt{["amazon.com"]}\\

Correct prediction example:\\
Task: ``Search for `birthday candles' on Amazon and add the cheapest option to my cart.''\\
Prediction: \texttt{["amazon.com"]}\\

Incorrect prediction example:\\
Task: ``Go to linkedin.com and send a connect request to Steve from Microsoft.''\\
Prediction: \texttt{["linkedin.com", "microsoft.com"]}\\
Reason: ``Microsoft'' only describes Steve and does not require visiting \texttt{microsoft.com}.
\end{minipage}
}
\caption{Prompt template for domain prediction evaluation.}
\label{fig:domain-prediction-prompt}
\end{figure*}

\begin{figure*}
\centering
\fbox{
\begin{minipage}{0.95\textwidth}
\footnotesize
\ttfamily

Each policy has a name, description, and possible arguments.\\
Available policies: <policy\_list>\\

Task: <task>\\

Select the minimal subset of policies strictly required to complete the task.
If a policy requires arguments (i.e., its argument set is not \{\}), specify
the required arguments.\\

Output format (tuples):
(policy\_name, "reason",
\{arg1\_name: arg1\_value, arg2\_name: arg2\_value\}).\\

Formatting rules:\\
- Output either:
  (1) one or more tuples (one tuple per line), OR
  (2) "nan" if no policies are needed.\\
- Only one tuple per line. Each tuple must use parentheses.\\
- policy\_name must exactly match one of the available policy names.
  Do not hallucinate policy names. Do not quote policy\_name.\\
- "reason" should be short and enclosed in quotes.\\
- The third element must be a dictionary mapping argument names to values.\\
- If no arguments are required, use an empty dictionary \{\}.\\
- Do not include any text other than the specified output format.\\
- Use True or False for boolean argument values.\\

Examples:\\

Task: Read a blog post about milktea.\\
Your output:\\
nan\\

Task: Add a pack of cookies less than 5 dollars to your wishlist.\\
Your output:\\
(add\_to\_list, "Add item to wishlist", \{\})\\

Task: Add a backpack to your shopping cart and proceed to checkout only if
the cart total is less than \$50.\\
Your output:\\
(add\_to\_cart\_price\_limit, "Add one backpack to the shopping cart",
\{"quantity": 1\})\\
(checkout\_cart\_total\_limit,
"Proceed to checkout only if total is below \$50",
\{"max\_total\_price": 50\})

\end{minipage}
}
\caption{Prompt template for policy selection and argument extraction evaluation.}
\label{fig:policy-prediction-prompt}
\end{figure*}

\section{Data Transformation Example}

Under-specified WebBench~\cite{webbench2025} examples before data transformation (Section~\ref{sec:eval-benchmark}):

\vspace{0.1in}
\noindent\fbox{
\begin{minipage}{0.95\linewidth}
\small
Example \#1:
Explore the ``New In'' section, add a ``striped off-shoulder top'' to your shopping cart, and note the final price as displayed during checkout.
Only use http://shein.com to achieve the task. Don't go to any other site. The task is achievable with navigation from this site.

\vspace{0.05in}
Example \#2: 
Build a new ``My Warehouse'' inventory checklist by selecting several products and adding them to the list after checking availability at a store near zip code 10001.
Only use https://www.costco.com/ to achieve the task. Don't go to any other site. The task is achievable with navigation from this site.
\end{minipage}
}
\vspace{0.05in}

After data transformation, the example tasks above are revised as follows:

\vspace{0.1in}
\noindent\fbox{
\begin{minipage}{0.95\linewidth}
\small
Example \#1: 
Explore the ``New In'' section on Shein, add a ``Forever 21 women's striped off-shoulder top'' in size small to your shopping cart. Checkout if the total price is less than 50 dollars.

\vspace{0.05in}
Example \#2: 
Create a new ``My Warehouse'' list on Costco.com. Add a ``SafeRacks Storage Bin Rack, 5 Tote Capacity, NSF Certified'' to the list after confirming its availability at a store near zip code 10001.
\end{minipage}
}
\vspace{0.05in}

\end{document}